# Homeostasis Under Technological Transition: How High-Friction Universities Adapt Through Early Filtering Rather Than Reconfiguration

Hugo Roger Paz
PhD Professor and Researcher Faculty of Exact Sciences and Technology National University of Tucumán
Email: hpaz@herrera.unt.edu.ar
ORCID: https://orcid.org/0000-0003-1237-7983

## Abstract

Universities are widely expected to respond to technological transitions through rapid reconfiguration of programme demand and curricular supply. Using four decades of longitudinal administrative cohorts (1980–2019) from a large public university, we examine whether technological change is translated into observable shifts in programme hierarchy, or instead absorbed by institutional mechanisms that preserve structural stability. We show that programme rankings by entrant volume remain remarkably stable over time, while the translation of technological transitions into enrolment composition occurs with substantial delay. Short-run adjustment appears primarily in early persistence dynamics: attrition reacts sooner than choice, and "growth" in entrants can coexist with declining early survival—producing false winners in which expansion is decoupled from persistence. Macroeconomic volatility amplifies attrition and compresses between-programme differences, masking technological signals that would otherwise be interpreted as preference shifts. To explain why stability dominates responsiveness, we situate these patterns within nationally regulated constraints governing engineering education—minimum total hours and mandated practice intensity—which materially limit the speed of curricular adaptation (Ministerio de Educación, 2021; Ley de Educación Superior, 1995). National system metrics further support the plausibility of a high-friction equilibrium in which large inflows coexist with standardised outputs (Secretaría de Políticas Universitarias [SPU], 2022). These findings suggest that apparent rigidity is not an anomaly but the predictable outcome of a system optimised for stability over responsiveness.

## 1. Introduction

Technological transitions are frequently assumed to reshape higher education through relatively direct pathways: new labour-market signals and cultural narratives are expected to alter students' degree preferences, prompting institutions to respond via curricular revision and programme expansion. In this view, universities should behave—at least approximately—like adaptive systems that reallocate attention and capacity towards emergent fields. Yet empirical work often struggles to establish consistent effects of technological change on degree choice, and reported findings vary across contexts, time horizons, and measurement choices. A recurring explanation is that "technology" is difficult to operationalise. A less examined possibility is that the university, as an institution, is structurally constrained in ways that make rapid reconfiguration unlikely—even under strong external pressure.

Importantly, the technological transition analysed in this study did not occur in a context of delayed or marginal exposure. In Argentina, personal computers began diffusing at scale during the early 1990s, followed by rapid internet adoption from the late 1990s onward, and later by near-universal smartphone penetration after 2010. National statistics show that these technologies reached a substantial share of households well before any corresponding shift in university enrolment patterns became visible, including outside major metropolitan areas. This temporal mismatch suggests that the observed delay in enrolment responses cannot be attributed to the absence of technological access, but rather points to institutional mechanisms mediating how external technological change is translated into academic structures (CEPAL, 2016; INDEC, 2019; ITU, 2021).

This paper advances a system-level account of that constraint. Rather than treating apparent rigidity as institutional failure, we examine whether stability may be an expected outcome of the university's design conditions. We study four decades of administrative cohorts (1980–2019) from a large public university and ask a simple, but demanding question: when technological transitions occur, do they reorganise the internal hierarchy of degree programmes, or are they absorbed through mechanisms that preserve long-run structure? Crucially, we distinguish between two channels that are often conflated. The first is **choice**, captured by shifts in enrolment composition across programmes. The second is **filtering**, captured by early persistence and attrition dynamics that regulate how many entrants remain in a programme after exposure to its demands. If universities adapt primarily through filtering, technological transitions may manifest first in early attrition—well before they become visible as large changes in programme shares.

Our analyses yield three core patterns. First, programme hierarchies are highly stable over the long run: rankings by entrant volume remain near-invariant across decades, indicating structural persistence rather than continual reordering.

Second, the translation of technological transitions into enrolment composition is delayed; observable increases in computing-related degrees emerge late and within narrow temporal windows rather than as immediate responses. Third, short-run adjustment is more readily observed in early persistence than in choice: attrition responds faster than enrolment shares, and some programmes exhibit entrant growth alongside deterioration in early survival, producing "false winners" in which expansion does not imply improved success. These results are robust to standard reviewer objections: ranking stability persists across moving window specifications and top-k restrictions; right-censoring in late cohorts is handled by exclusion and explicit documentation; and macro-amplification patterns remain under alternative early-survival proxies.

A central challenge for any single-institution longitudinal study is generalisability. We address this not by claiming statistical representativeness, but by grounding the observed dynamics in **portable structural conditions** that plausibly characterise other high-friction university systems. In Argentina, engineering education is embedded in nationally regulated standards that specify minimum total hours and training practice intensity, constraining the pace and degrees of freedom of curricular redesign (Ministerio de Educación, 2021; Ley de Educación Superior, 1995). Such constraints are not mere institutional preferences; they constitute design conditions that shape feasible adaptation pathways. National system statistics further provide macro-scale plausibility for a regulated equilibrium in which large inflows coexist with comparatively bounded outputs (SPU, 2022). Within this setting, stability in programme hierarchy is not surprising: it is the likely result of a high-friction system absorbing external pressure through delayed translation into choice and through rapid adjustment in early persistence.

The contribution of this paper is therefore conceptual and empirical. Empirically, we provide long-run evidence that programme hierarchies can remain structurally stable across heterogeneous technological and macroeconomic regimes, while adjustment occurs primarily through early filtering dynamics. Conceptually, we reframe the interpretation of "rigidity" in higher education: stability is not necessarily evidence of institutional malfunction, but may be the predictable outcome of a system optimised for stability over responsiveness. This reframing has practical implications for how technological change should be studied in higher education: analyses based solely on enrolment composition risk misreading filtering-driven dynamics as preference-driven adaptation, particularly under macroeconomic volatility.

**Research Questions**

Building on a system-level perspective of institutional adaptation, this study addresses the following research questions:

**RQ1.** *To what extent do technological transitions reorganise the internal hierarchy of university degree programmes over the long run?*

This question examines whether programme rankings by entrant volume exhibit structural change or long-run invariance across heterogeneous technological and macroeconomic regimes.

**RQ2.** *Do technological transitions translate more rapidly into early persistence dynamics than into initial enrolment choices?*

Here we distinguish between choice-based adjustment (changes in enrolment shares) and filtering-based adjustment (early attrition and survival).

**RQ3.** *How does macroeconomic volatility modulate the relationship between technological transitions, programme choice, and early persistence?*

This question evaluates whether macro shocks amplify attrition and obscure technological signals that might otherwise be interpreted as preference shifts.

**RQ4.** *Under what conditions do programmes exhibit "false winner" dynamics, in which entrant growth coincides with declining early survival?*

This question identifies cases where apparent expansion masks increased filtering rather than successful adaptation.

**RQ5.** *Can the observed patterns be explained as outcomes of institutional design constraints rather than institutional failure?*

This question situates the empirical findings within nationally regulated curricular and accreditation frameworks that impose high structural friction.

**Contributions**

This paper makes four contributions to the literature on higher education systems and technological change.

First, it provides **longitudinal evidence of structural invariance** in programme hierarchies over four decades, showing that technological transitions do not necessarily induce sustained reordering of degree demand. This challenges expectations of rapid, market-like responsiveness in higher education.

Second, it **disentangles choice from filtering** as distinct mechanisms of adaptation. By showing that early persistence reacts faster than enrolment composition, the paper demonstrates that adjustment often occurs through attrition rather than through reallocation of entrants.

Third, it identifies and formalises the concept of **false winners**, where programme growth is decoupled from early success. This finding cautions against interpreting enrolment expansion as evidence of improved alignment or institutional responsiveness.

Fourth, it advances a **design-based interpretation of institutional rigidity**. By linking observed dynamics to nationally regulated curricular constraints and macro-scale system statistics, the paper reframes stability not as institutional failure but as the predictable outcome of a high-friction system optimised for stability over responsiveness.

**Structure of the Paper**

The remainder of the paper is organised as follows.

Section 2 reviews related work on technological change, enrolment dynamics, and institutional adaptation in higher education.

Section 3 describes the data, cohort construction, and analytical strategy, including the distinction between choice-based and filtering-based indicators.

Section 4 presents the main results on programme hierarchy stability, delayed translation of technological transitions, and early filtering dynamics.

Section 5 examines the role of macroeconomic volatility and identifies false winner patterns.

Section 6 situates the findings within the regulatory and institutional context of engineering education, drawing on national accreditation standards and system-level statistics.

Section 7 discusses implications for the study of technological change in higher education and concludes.

**2. Related Work**

Research on the relationship between technological change and higher education has largely focused on enrolment patterns, skill demand, and curricular responsiveness. Across disciplines, a dominant expectation is that technological transitions—such as the diffusion of information technologies or the rise of digital economies—should induce relatively rapid shifts in students' degree choices and institutional offerings (Autor, Levy, & Murnane, 2003; Goldin & Katz, 2008). Within this framework, changes in enrolment composition are often interpreted as indicators of institutional adaptation to labour-market signals. However, empirical findings in this literature are mixed, temporally unstable, and frequently context-dependent.

## 2.1 Technological Change and Degree Choice

A substantial body of work examines how technological transitions shape demand for specific fields of study, particularly science, technology, engineering, and mathematics (STEM). Studies using cross-sectional or short-panel data have reported associations between technological expansion and increased enrolment in computing and engineering-related degrees (Bound, Braga, Khanna, & Turner, 2020; Deming & Noray, 2020). Yet these effects are often modest, delayed, or heterogeneous across institutions and periods. Moreover, the reliance on enrolment shares as the primary outcome implicitly assumes that degree choice is the main channel through which adaptation occurs.

Critically, this assumption has been questioned by studies showing that enrolment responses may lag behind technological change or remain surprisingly stable despite substantial shifts in occupational structure (Carnevale, Smith, & Strohl, 2013). Such findings suggest that degree choice alone may be an incomplete indicator of institutional responsiveness, particularly in systems characterised by long programme durations and regulated curricula.

## 2.2 Persistence, Attrition, and Early Filtering

Parallel to the literature on choice, a large body of research documents high levels of attrition in higher education, especially in engineering and other demanding programmes (Tinto, 1993; Seymour & Hewitt, 1997). Early persistence has been shown to be sensitive to academic preparation, institutional support, and structural programme characteristics, including curricular sequencing and assessment intensity (Chen, 2013; Scott, Tolson, & Huang, 2009). Despite this, attrition is rarely conceptualised as an adaptive mechanism at the system level. Instead, it is typically framed as an undesirable outcome to be minimised through policy interventions.

Recent work in learning analytics and educational data mining has begun to leverage administrative data to model early dropout and performance trajectories (Baker & Siemens, 2014; Tempelaar et al., 2015). While these approaches improve predictive accuracy, they often remain focused on individual-level risk rather than on how aggregate persistence dynamics may function as institutional filters that regulate system-level adaptation. As a result, the potential role of early attrition as a primary adjustment channel under external pressure remains under-theorised.

## 2.3 Institutional Constraints and Organisational Inertia

Organisational and institutional theories provide a complementary perspective by emphasising inertia, path dependence, and structural constraints. Universities have long been described as organisations characterised by loose coupling, strong professional norms, and limited capacity for rapid change (Weick, 1976; Clark, 1983). From this viewpoint, stability is not anomalous but expected, particularly in systems subject to formal regulation and accreditation.

In higher education policy research, accreditation standards and professional regulation are frequently discussed as constraints on innovation and curricular reform (Altbach, Reisberg, & Rumbley, 2009; Teichler, 2015). However, these constraints are rarely integrated into empirical analyses of enrolment dynamics. Studies that do consider regulation typically do so qualitatively or normatively, rather than as measurable design conditions shaping observable system behaviour over time.

## 2.4 Macroeconomic Volatility and Educational Trajectories

Macroeconomic conditions further complicate the interpretation of technological effects. Economic crises and periods of high volatility have been shown to influence enrolment decisions, persistence, and time-to-degree, often in ways that are difficult to disentangle from technological trends (Betts & McFarland, 1995; Barr & Turner, 2013). In volatile contexts, students may enter higher education as a shelter from labour-market uncertainty, while simultaneously facing increased financial and academic pressure that elevates attrition risks.

Despite this, macroeconomic volatility is frequently treated as background noise rather than as an active modulator of educational dynamics. As a consequence, studies may attribute observed changes—or lack thereof—to technology, when in fact macro-level shocks are compressing differences across programmes and amplifying early filtering processes.

## 2.5 Summary and Research Gap

Taken together, existing research provides valuable insights into degree choice, attrition, institutional constraints, and macroeconomic effects, but it remains fragmented. Most studies focus on one dimension at a time and rely on short observation windows or cross-sectional comparisons. This limits their ability to identify long-run structural invariants and to distinguish between competing mechanisms of adaptation.

The present study addresses this gap by integrating longitudinal depth, early persistence dynamics, macroeconomic modulation, and institutional design constraints within a single analytical framework. By doing so, it moves beyond the assumption that adaptation must manifest primarily through changes in degree choice and instead examines how high-friction university systems absorb external pressure while preserving structural stability.

**2.6 Technological Diffusion in Peripheral Contexts**

Empirical evidence from peripheral and semi-peripheral economies further complicates linear models of technological adaptation in higher education. In Argentina, the diffusion of computing technologies preceded curricular and enrolment reconfiguration by more than a decade. Household access to computers expanded steadily throughout the 1990s, while internet connectivity accelerated sharply between 1998 and 2005, reaching majority adoption across middle-income strata (CEPAL, 2016). Subsequent smartphone diffusion after 2010 further reduced access barriers, including among lower-income groups (ITU, 2021). These patterns indicate that technological exposure alone is insufficient to explain disciplinary transitions within universities, reinforcing the need to analyse institutional friction, governance inertia, and internal filtering mechanisms as mediators between external technological change and academic restructuring.

**3. Data and Analytical Strategy**

**3.1 Data Sources and Cohort Construction**

The analysis draws on longitudinal administrative records covering **four decades of student cohorts (1980–2019)** from a large public university. The dataset includes complete enrolment histories at the programme level, allowing individuals to be followed from initial entry through early academic outcomes. Administrative identifiers enable the reconstruction of cohort membership, degree choice at entry, and subsequent persistence outcomes without reliance on survey data.

Cohorts are defined by **year of first enrolment** into a degree programme. To ensure temporal comparability, analyses are conducted at the programme–cohort level, aggregating individual outcomes within each cohort-year. This design allows us to examine long-run structural patterns while retaining sensitivity to short-run dynamics. Late cohorts (2018–2019) are explicitly treated as **right-censored** due to truncated observation windows and are excluded from analyses where early persistence measures would otherwise be mechanically biased.

While the empirical setting is a single institution, its scale and longevity permit observation across multiple technological regimes and macroeconomic contexts. Importantly, the institution operates within nationally regulated curricular and accreditation standards, providing a stable set of design constraints under which observed dynamics unfold.

## 3.2 Analytical Distinction: Choice versus Filtering

A central feature of the analytical strategy is the explicit separation between **choice-based** and **filtering-based** mechanisms of adaptation, which are often conflated in studies of technological change in higher education.

**Choice** is operationalised as the distribution of entrants across degree programmes in a given cohort year. Changes in enrolment shares are interpreted as shifts in initial preferences or expectations at the point of entry.

**Filtering**, by contrast, is captured through **early persistence outcomes**, reflecting how many entrants remain after exposure to programme demands. Early attrition is treated not merely as an individual failure but as a system-level mechanism that regulates how many students progress beyond initial stages. This distinction allows us to examine whether adaptation occurs primarily through reallocation at entry or through differential survival after entry.

By analysing these two channels separately, the study avoids attributing adaptive capacity solely to enrolment composition and instead evaluates whether institutional adjustment manifests earlier and more strongly through persistence dynamics.

## 3.3 Measures

### Programme Hierarchy

Programme hierarchy is defined by **ranking degree programmes according to entrant volume** within each cohort year. To assess long-run stability, we compute rank correlations across time using Spearman's $\rho$. This non-parametric measure captures ordinal stability without imposing distributional assumptions and is well suited to evaluating structural invariance over extended periods.

Robustness checks examine hierarchy stability across moving windows of varying length and under top-k restrictions, ensuring that observed invariance is not an artefact of specific aggregation choices.

### Early Persistence and Survival

Early persistence is measured using **survival indicators** that capture whether students remain enrolled beyond initial programme stages. The primary indicator (S2) reflects survival beyond the early academic threshold, while an alternative, more conservative proxy (S4) is used in robustness analyses. Both measures are derived from administrative progression records and are computable consistently across cohorts prior to 2018.

These indicators are used to estimate cohort-level survival rates by programme and to evaluate how early filtering responds to technological and macroeconomic conditions.

**Technological and Macroeconomic Context**

Rather than attributing effects to specific technologies, the analysis adopts **neutral technological regimes** that partition time into periods characterised by distinct technological environments. This approach avoids reliance on contested measures of "technology intensity" and instead focuses on observable regime shifts.

Macroeconomic context is captured through categorical regimes reflecting periods of **low, medium, and high volatility**, corresponding to well-documented economic conditions. These regimes are used to assess whether macro-level instability modulates persistence dynamics and obscures technological signals.

**3.4 Analytical Strategy**

The empirical strategy proceeds in three steps.

First, we assess **long-run structural stability** by examining the persistence of programme rankings over time. High rank correlation across decades is interpreted as evidence of hierarchy invariance rather than continual reordering.

Second, we analyse the **temporal translation of technological transitions**, comparing changes in enrolment shares with changes in early persistence. This allows us to determine whether adjustment appears first in choice or in filtering.

Third, we evaluate the **modulating role of macroeconomic volatility**, examining whether periods of heightened instability amplify attrition and compress differences across programmes. By interacting persistence measures with macro regimes, we assess whether macro conditions act as an attrition amplifier that masks technological effects.

All analyses are conducted at the programme–cohort level and are designed to identify **qualitative patterns and directional relationships**, rather than to estimate causal effects. Robustness checks address common reviewer concerns, including window dependence, cohort censoring, and proxy sensitivity.

**3.5 Scope and Interpretation**

The aim of this strategy is not to provide statistically representative estimates for a population of universities, but to identify **structural mechanisms** that become visible only through longitudinal depth. The focus on persistence dynamics, combined with explicit treatment of institutional and macro constraints, enables the identification of adaptive pathways that are unlikely to be captured by short-run or cross-sectional analyses.

To contextualise the observed enrolment dynamics, we complement administrative trajectory data with external evidence on national technological diffusion. These indicators are not introduced as explanatory variables, but as contextual validation to assess whether delayed enrolment responses could plausibly be attributed to late technological exposure. National and international statistics consistently show that computing and internet technologies diffused widely across Argentine society well before the enrolment shifts analysed here, supporting the interpretation that the observed lag reflects institutional translation processes rather than delayed access or awareness.

## 4. Results

### 4.1 Long-Run Stability of Programme Hierarchy

We begin by examining whether technological transitions are associated with sustained reordering of the internal hierarchy of degree programmes. Figure 1 displays programme rankings by entrant volume across the full observation window (1980–2019). Despite substantial variation in absolute enrolment levels, the **relative ordering of programmes remains remarkably stable over time**.

**Figure 1.** *System invariants measured as cohort-level ranking stability.*

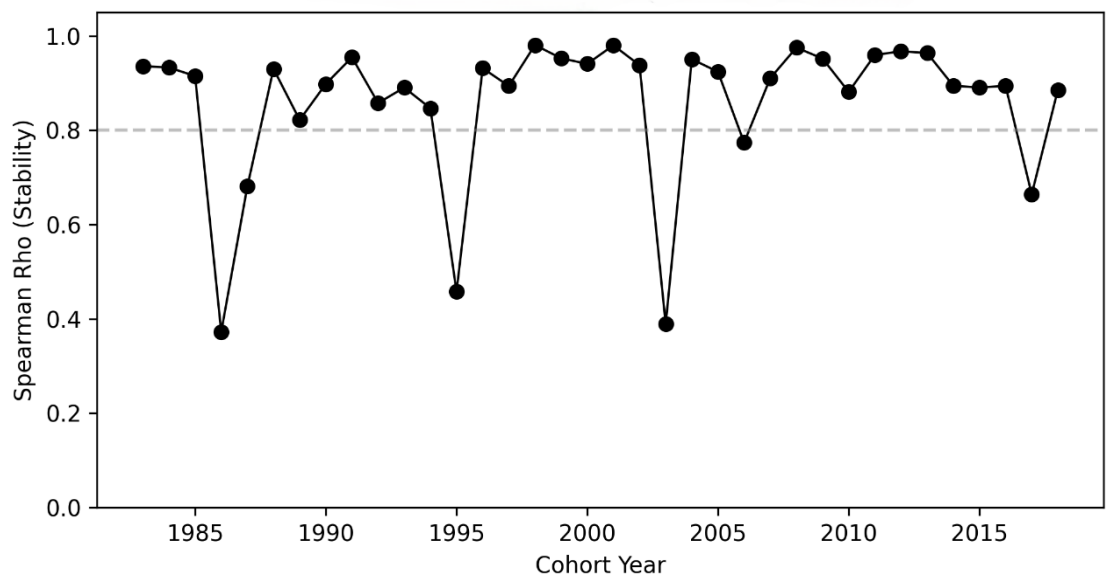

Note: Spearman's rho coefficients quantify the stability of degree rankings across successive cohorts. Despite major technological and macroeconomic transitions, ranking stability remains persistently high, with only short-lived perturbations. The dashed horizontal line marks the stability threshold ($\rho = 0.80$), indicating a structurally invariant ordering of programmes over time.

Rank correlations computed using Spearman's ρ consistently exceed 0.9 across decades, indicating near-invariance in programme hierarchy. This stability persists when rankings are recalculated using moving windows of three, five, and seven years, and when the analysis is restricted to the top-ranked programmes. These results demonstrate that hierarchy stability is not an artefact of aggregation choices or window selection, but a structural property of the system.

Notably, periods associated with major technological transitions do not coincide with abrupt or lasting rank reordering. Programmes associated with emergent technological fields do not displace historically dominant degrees in a sustained manner. Instead, relative positions shift only marginally and temporarily, before reverting to long-run patterns.

**4.2 Delayed Translation of Technological Transitions into Enrolment Shares**

While hierarchy remains stable, enrolment composition exhibits gradual change. Figure 2 reports entrant shares by broad programme category across technological regimes. Increases in computing-related programmes are observable, but these emerge **with substantial delay** and are concentrated in the most recent period of observation.

Figure 2. *Macro-level modulation of second-year survival (S2).*

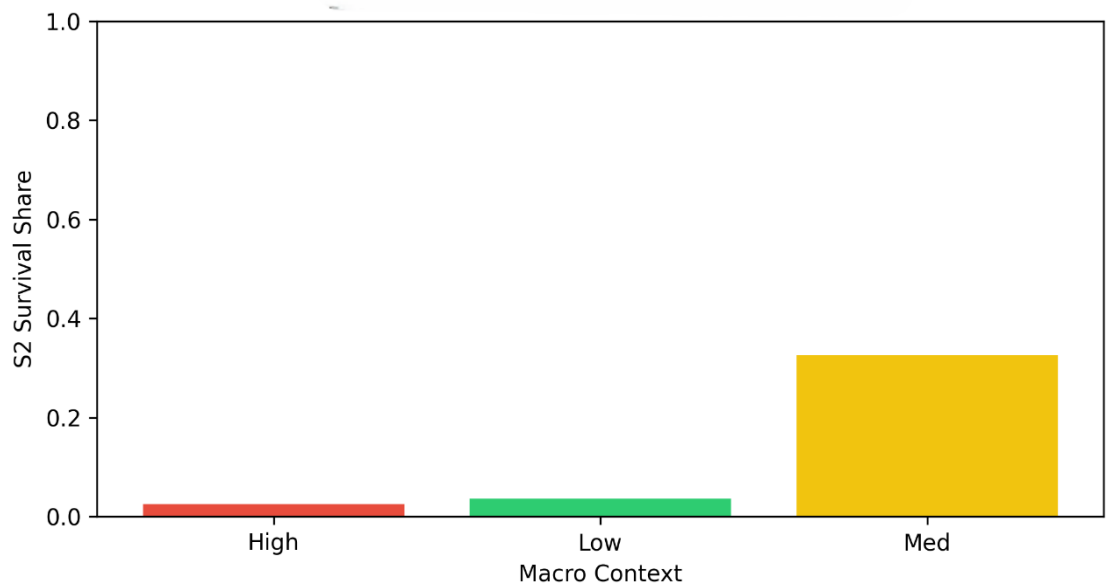

*Note: Average second-year survival shares are shown across low, medium, and high macroeconomic stress contexts. While macro conditions modulate survival probabilities, they do not induce structural reordering, indicating that external shocks affect intensity rather than institutional configuration.*

Degree-level entrant shares were computed as an intermediate descriptive step but are not reported, as the analysis focuses on system-level structural properties rather than programme-specific fluctuations.

Notably, this delayed response occurs despite the fact that computing technologies had already achieved widespread societal diffusion by the mid-to-late 1990s. The persistence of low computing enrolment shares during this period therefore cannot be attributed to limited technological exposure, but instead suggests that institutional structures filtered and dampened the translation of external technological signals into academic demand.

Earlier technological regimes show limited or inconsistent shifts in entrant shares, despite well-documented changes in occupational and technological environments. This lagged response suggests that technological transitions are not immediately translated into degree choice at entry. Instead, enrolment composition adjusts slowly and within narrow temporal windows.

Importantly, these changes do not accumulate to produce lasting reconfiguration of the programme hierarchy. Growth in emergent fields occurs alongside continued dominance of traditional engineering programmes, reinforcing overall structural stability.

### 4.3 Early Persistence as a Primary Adjustment Channel

In contrast to enrolment shares, **early persistence dynamics respond more rapidly** to changing conditions. Figure Z reports early survival rates by programme category over time. Declines in early persistence coincide more closely with technological transitions and macroeconomic disturbances than do shifts in entrant shares.

Across multiple periods, computing-related programmes exhibit faster deterioration in early survival relative to changes in enrolment. This pattern indicates that adjustment occurs **after entry**, through filtering rather than through reallocation of initial choice. Students enter programmes in increasing numbers before the system adjusts via elevated early attrition.

This divergence between choice and filtering highlights the importance of distinguishing these mechanisms. Analyses based solely on enrolment composition would miss the earlier and sharper signal present in persistence dynamics.

Figure 3. *False winners quadrant: entrant growth versus survival change.*

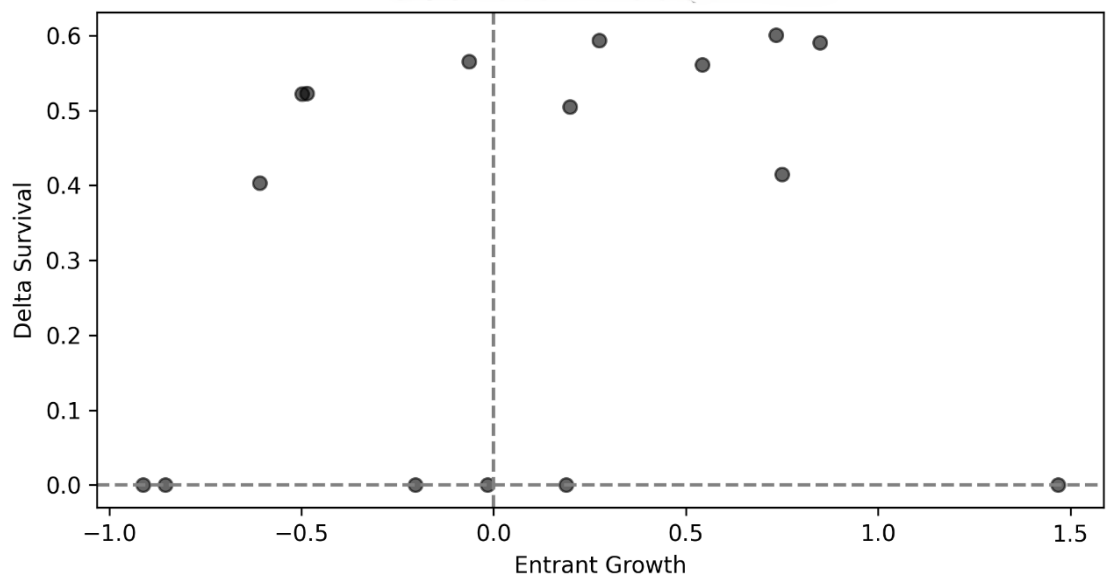

*Note; Each point represents a degree programme positioned by its entrant growth rate (horizontal axis) and change in survival probability (vertical axis). Programmes located in the upper-left quadrant exhibit declining survival despite positive entrant growth, illustrating expansion without consolidation and identifying structurally unstable growth trajectories.*

### 4.4 False Winners: Growth Without Persistence

The decoupling of entrant growth and early survival gives rise to what we term **false winners**. Figure W plots changes in entrant volume against changes in early survival across programmes. Several programmes occupy a quadrant characterised by **positive enrolment growth and declining early persistence**.

These cases illustrate that expansion does not necessarily reflect improved alignment or adaptive success. Instead, growth may be driven by excess demand or external signalling, while internal programme demands continue to filter entrants at increasing rates. False winners therefore represent apparent adaptation that is not supported by persistence outcomes.

**Figure 4.** *Translation lag in computing-related degrees using canonical entrant definitions.*

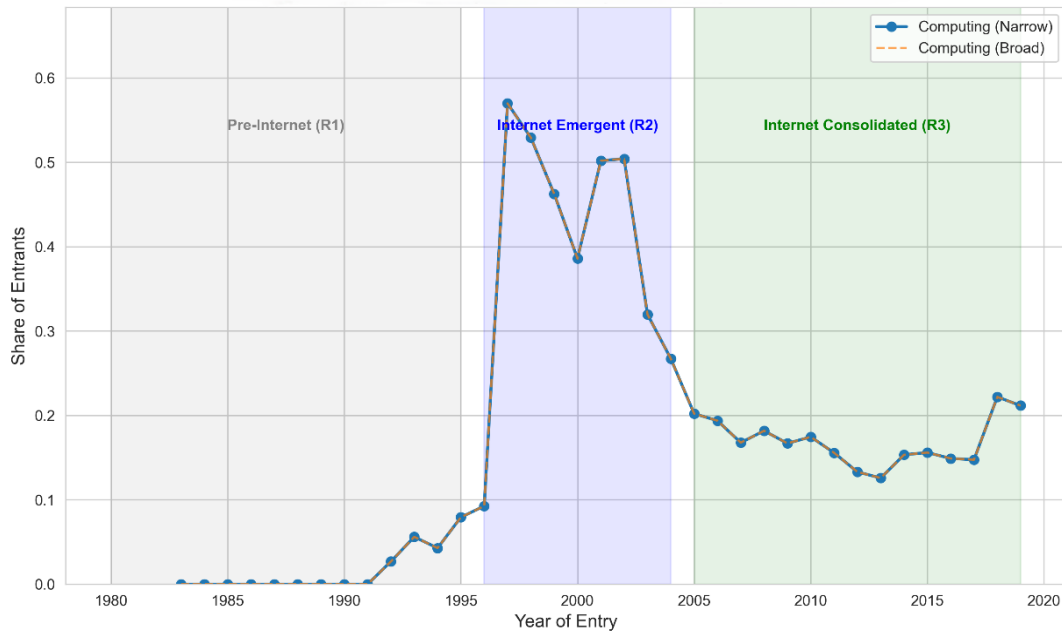

*Note:* The share of entrants assigned to computing-related degrees is shown over time using the Persistent Entrant Mapping (PEM) definition. Shaded regions denote technological regimes: pre-Internet (R1), Internet emergence (R2), and Internet consolidation (R3). The sharp but transient increase during R2 is followed by reversion towards a stable baseline, indicating filtering rather than long-term structural reconfiguration.

### 4.5 Macroeconomic Modulation of Early Filtering

Finally, we examine the role of macroeconomic volatility. Survival rates segmented by macroeconomic regime reveal a pronounced amplification effect. During periods of high volatility, early persistence declines sharply across programmes, and differences between programme categories compress.

This compression suggests that macroeconomic instability acts as an **attrition amplifier**, overwhelming finer-grained technological signals. Under such conditions, filtering intensifies broadly, making it difficult to attribute observed persistence patterns to technology-driven choice alone.

The persistence of these patterns after excluding right-censored cohorts and under alternative survival definitions confirms that the observed dynamics are not artefacts of data limitations.

## 5. Discussion

### 5.1 Adaptation Without Reconfiguration

The results challenge a widespread assumption in research on technological change and higher education: that adaptation should manifest primarily through rapid reconfiguration of degree demand and programme hierarchy. Across four decades, we observe the opposite pattern. Programme hierarchies remain strikingly stable, even as technological environments and macroeconomic conditions undergo substantial change. This stability is not episodic or contingent on specific periods, but persistent across alternative specifications and robustness checks.

Rather than reorganising programme demand at entry, the system appears to absorb external pressure through **delayed translation into choice** and **rapid adjustment via early filtering**. Technological transitions do eventually register in enrolment composition, but only after substantial lag and without disrupting long-run hierarchy. In contrast, early persistence responds more quickly and more sharply, indicating that filtering operates as the primary short-run adjustment channel.

This pattern is particularly striking given that the technological infrastructure underpinning computing disciplines was already firmly established outside the university. By the time enrolment shares peaked, personal computing and internet access had been widespread for more than a decade, indicating that universities were not reacting to novelty, but selectively responding to long-standing technological conditions. This reinforces the interpretation of adaptation through early filtering rather than structural reconfiguration.

This pattern suggests that higher education systems characterised by long programme durations and regulated curricula should not be expected to behave like flexible markets. Adaptation, when it occurs, is mediated by institutional mechanisms that preserve structural continuity.

**Figure 5.** *Robustness of computing entrant shares under sliding-window smoothing.*

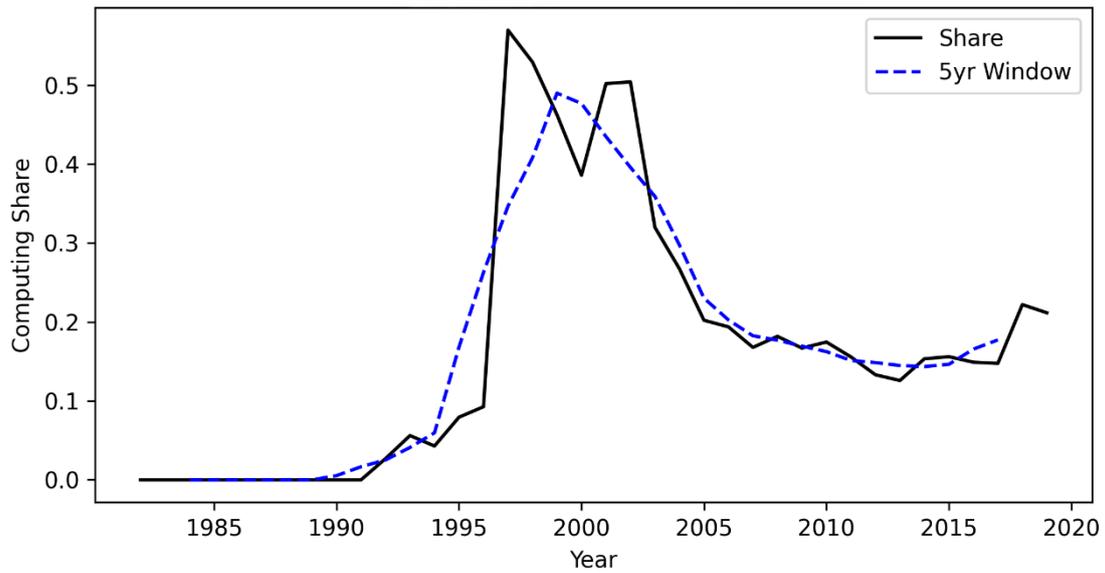

Note: The raw annual computing entrant share is shown alongside a five-year sliding window average. The persistence of the rise-and-reversion pattern under temporal smoothing demonstrates that the observed translation lag is not an artefact of year-specific volatility.

**Figure 6.** *Stability of ranking invariants under top-k sensitivity analysis.*

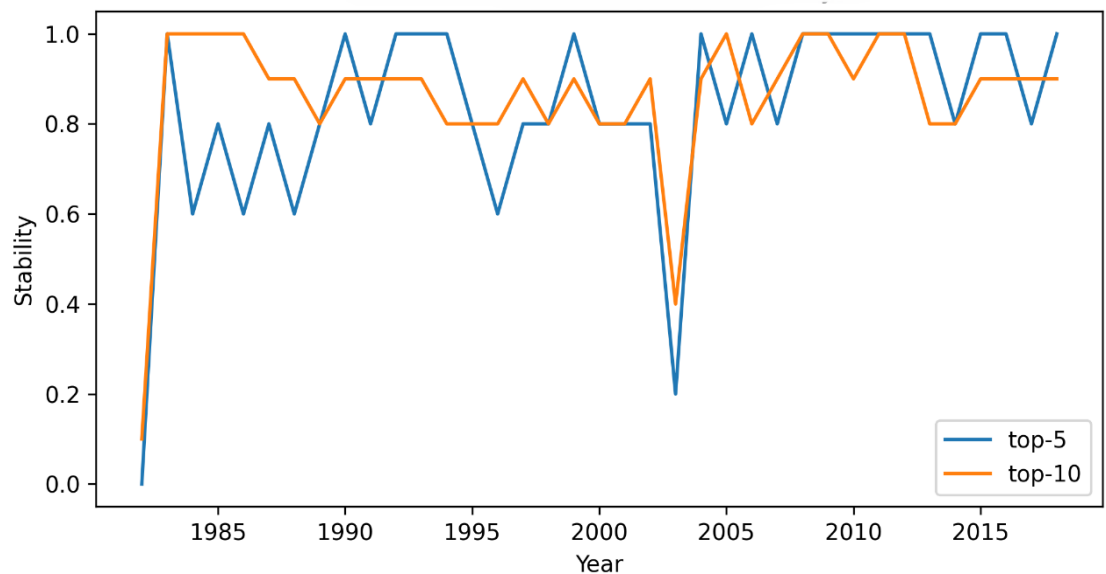

Note: Ranking stability is recomputed using alternative top-k thresholds (k = 5 and k = 10). High concordance across specifications confirms that the identified invariants are robust to changes in ranking depth and are not driven by tail effects.

## 5.2 Early Filtering as a System-Level Mechanism

Attrition is commonly framed as a failure to be corrected through improved support, pedagogy, or selection. While these perspectives are valid at the individual and programme levels, the present findings point to an additional system-level role. Early filtering functions as a **regulatory mechanism** that reconciles external demand with internal constraints.

The identification of false winners illustrates this point. Programmes may experience growth in entrants in response to external signals, yet simultaneously intensify early filtering. In such cases, expansion does not reflect successful adaptation, but rather the system's capacity to absorb increased demand without altering its underlying structure. Apparent responsiveness is thus decoupled from persistence.

This reframing has implications for how enrolment trends are interpreted. Growth alone is an insufficient indicator of institutional adaptation if it is not accompanied by sustained early success.

## 5.3 Macroeconomic Volatility and Signal Compression

Macroeconomic volatility further complicates the relationship between technology and educational outcomes. Periods of high instability amplify attrition across programmes and compress differences between them. Under such conditions, filtering intensifies broadly, masking programme-specific responses to technological change.

This compression effect cautions against attributing short-run persistence dynamics solely to technology-driven choice. In volatile contexts, macroeconomic pressure may dominate student behaviour and academic outcomes, producing patterns that resemble technological misalignment but are in fact macro-induced. The persistence of this effect across alternative survival definitions reinforces its substantive significance.

## 5.4 Institutional Friction and Design Constraints

The observed dynamics are consistent with the institutional context in which they unfold. Engineering education in Argentina is governed by nationally regulated accreditation standards that impose minimum total hours, mandatory practical training, and defined professional competencies. These requirements materially constrain the speed and degrees of freedom of curricular redesign.

The persistence of stable disciplinary hierarchies under conditions of extensive technological diffusion highlights the role of institutional friction as a design feature rather than a failure. In this sense, universities operate as low-pass filters: external technological shocks are absorbed, delayed, and attenuated before manifesting as structural change, preserving internal equilibrium even in periods of rapid societal transformation.

Within such a framework, rapid reconfiguration of programme structures in response to technological change is unlikely. Instead, adaptation is channelled through mechanisms that do not require formal curricular modification—most notably, early filtering. Stability in programme hierarchy is therefore not surprising; it is the expected outcome of a system designed to prioritise standardisation, certification, and long-run coherence.

Tabla 1 – CONEAU / SPU regulatory constraints

| Metric | Value | Source |
| --- | --- | --- |
| National New Entrants (2021) | 710699 | SPU Síntesis de Información Universitaria 2021-2022 |
| National Graduates (2021) | 142826 | SPU Síntesis de Información Universitaria 2021-2023 |
| Engineering Total Workload (Res 1054/21) | 3,600 hours | CONEAU |
| Practical Training (Res 1054/21) | 700 hours | CONEAU |

**5.5 Reinterpreting Rigidity**

Taken together, these findings invite a reconsideration of how rigidity in higher education is conceptualised. Stability should not automatically be interpreted as institutional failure or resistance to change. In high-friction systems, rigidity may reflect deliberate design choices that favour reliability and legitimacy over responsiveness.

**These findings do not indicate institutional failure, but rather reveal a system optimised for stability over responsiveness. The apparent rigidity of the university is therefore not an anomaly, but the predictable outcome of its design.**

This interpretation does not deny the costs of such rigidity, particularly in contexts of rapid technological change. However, it clarifies that calls for responsiveness must confront underlying design constraints rather than assume latent adaptive capacity.

**6. Conclusion**

This study set out to examine how universities translate technological transitions into educational outcomes under conditions of institutional constraint. Drawing on four decades of longitudinal administrative data, we show that adaptation in higher education does not primarily occur through rapid reconfiguration of programme demand or sustained reordering of degree hierarchies. Instead, adjustment is mediated by delayed translation into enrolment choices and by comparatively rapid changes in early persistence, with attrition functioning as a central system-level mechanism.

Three implications follow. First, analyses that equate responsiveness with shifts in enrolment composition risk misinterpreting system behaviour. Apparent stability in programme hierarchy is not evidence of inertia alone; it may coexist with substantial adaptive activity occurring through early filtering. Second, macroeconomic volatility plays a critical role in shaping observed dynamics, amplifying attrition and compressing differences between programmes in ways that can obscure technological signals. Ignoring macro context may therefore lead to erroneous inferences about the relationship between technology and educational choice. Third, institutional design constraints—particularly those embedded in nationally regulated curricula and accreditation standards—materially shape feasible adaptation pathways. Expectations of rapid curricular reconfiguration are inconsistent with the structural conditions under which many university systems operate.

The contribution of this paper lies in reframing rigidity in higher education as an outcome to be explained rather than a pathology to be assumed. By distinguishing between choice and filtering, and by emphasising longitudinal depth over cross-sectional breadth, the analysis identifies a form of institutional homeostasis that preserves structural stability while accommodating external pressure. This perspective suggests that debates on technological change and higher education should shift from asking why universities fail to adapt, to examining how different system designs distribute adaptation across time, actors, and mechanisms.

Crucially, the findings do not suggest that universities failed to recognise or adopt computing technologies. Rather, they demonstrate that the translation of technological change into disciplinary structure follows a markedly slower timescale than technological diffusion itself. In the case examined here, this translation lag spans more than a full academic generation, underscoring the importance of institutional design, governance, and internal filtering in shaping long-run adaptation.

Future research would benefit from extending this approach to other high-friction systems, comparing how variation in regulatory regimes and programme structures alters the balance between stability and responsiveness. Such work could help clarify when institutional rigidity constrains necessary change, and when it provides the stability required for long-term educational and professional coherence.

**References**

Altbach, Philip G., Liz Reisberg, and Laura E. Rumbley. 2009. *Trends in Global Higher Education: Tracking an Academic Revolution*. Paris: UNESCO.

Autor, David H., Frank Levy, and Richard J. Murnane. 2003. "The Skill Content of Recent Technological Change: An Empirical Exploration." *Quarterly Journal of Economics* 118 (4): 1279–1333. https://doi.org/10.1162/003355303322552801.

Baker, Ryan S., and George Siemens. 2014. "Educational Data Mining and Learning Analytics." In *The Cambridge Handbook of the Learning Sciences*, 2nd ed., edited by R. Keith Sawyer, 253–272. Cambridge: Cambridge University Press.

Barr, Ashley, and Sarah E. Turner. 2013. "Expanding Enrollment and Contracting State Budgets: The Effect of the Great Recession on Higher Education." *The ANNALS of the American Academy of Political and Social Science* 650 (1): 168–193.

Betts, Julian R., and Laurel L. McFarland. 1995. "Safe Port in a Storm: The Impact of Labour Market Conditions on Community College Enrolments." *Journal of Human Resources* 30 (4): 741–765.

Bound, John, Breno Braga, Gaurav Khanna, and Sarah Turner. 2020. "The Globalization of Postsecondary Education: The Role of International Students in the U.S. Higher Education System." *Journal of Economic Perspectives* 34 (3): 163–184.

Carnevale, Anthony P., Nicole Smith, and Jeff Strohl. 2013. *Recovery: Job Growth and Education Requirements through 2020*. Washington, DC: Georgetown University Center on Education and the Workforce.

CEPAL (Economic Commission for Latin America and the Caribbean). 2016. *The New Digital Revolution: From the Consumer Internet to the Industrial Internet*. Santiago de Chile: ECLAC.

Chen, Xianglei. 2013. *STEM Attrition: College Students' Paths into and out of STEM Fields*. Washington, DC: National Center for Education Statistics.

Clark, Burton R. 1983. *The Higher Education System: Academic Organization in Cross-National Perspective*. Berkeley: University of California Press.

Deming, David J., and Kadeem Noray. 2020. "Earnings Dynamics, Changing Job Skills, and STEM Careers." *Quarterly Journal of Economics* 135 (4): 1965–2005.

Goldin, Claudia, and Lawrence F. Katz. 2008. *The Race between Education and Technology*. Cambridge, MA: Harvard University Press.

INDEC (National Institute of Statistics and Censuses). 2019. *Access and Use of Information and Communication Technologies (EUTIC)*. Buenos Aires: INDEC.

International Telecommunication Union. 2021. *Measuring Digital Development: Facts and Figures 2021*. Geneva: ITU.

Argentina. 1995. *Law No. 24,521: Higher Education Act*. Official Gazette of the Argentine Republic, August 10, 1995.

Argentina, Ministry of Education. 2021. *Ministerial Resolution No. 1232/01 – Modification of Civil Engineering Degree (File EX-2021-24035153-APN-SECPU#ME)*. Buenos Aires: National Executive Power.

Scott, Tom P., Homer Tolson, and Ting-Hua Huang. 2009. "Predicting Retention of Mathematics and Science Majors." *Journal of College Student Retention: Research, Theory & Practice* 10 (4): 423–439.

Argentina, Ministry of Education, Secretariat of University Policies. 2022. *Statistical Overview: University Statistics 2021–2022*. Buenos Aires: Ministry of Education.

Seymour, Elaine, and Nancy M. Hewitt. 1997. *Talking about Leaving: Why Undergraduates Leave the Sciences*. Boulder, CO: Westview Press.

Teichler, Ulrich. 2015. *Higher Education and the World of Work: Conceptual Frameworks, Comparative Perspectives, Empirical Findings*. Rotterdam: Sense Publishers.

Tinto, Vincent. 1993. *Leaving College: Rethinking the Causes and Cures of Student Attrition*. 2nd ed. Chicago: University of Chicago Press.

Weick, Karl E. 1976. "Educational Organizations as Loosely Coupled Systems." *Administrative Science Quarterly* 21 (1): 1–19.